\documentclass[10pt,aps,pre,showpacs,twocolumn]{revtex4-1}

\usepackage{graphicx}

\begin{document}



\title{Avalanches, Breathers and Flow Reversal in a Continuous Lorenz-96 Model}


\author{R. Blender}
\affiliation{Meteorologisches Institut, KlimaCampus, 
Universit\"at Hamburg, Hamburg, Germany}
\author{J. Wouters}
\author{V. Lucarini}
\affiliation{Meteorologisches Institut, KlimaCampus, 
Universit\"at Hamburg, Hamburg, Germany}


\date{\today}

\begin{abstract}
For the discrete model suggested by Lorenz in 1996 
a one-dimensional long wave approximation 
with nonlinear excitation and diffusion is derived.
The model 
is energy conserving but non-Hamiltonian.
In a low order truncation weak external forcing 
of the zonal mean flow induces 
avalanche-like breather solutions 
which cause reversal of the mean flow by a 
wave-mean flow interaction.
The mechanism is an outburst-recharge 
process similar to avalanches in a sand pile model. 
\end{abstract}


\pacs{47.20.-k   
      47.10.Df   
      47.35.Bb   
      64.60.Ht   
      }   


\maketitle


In 1996 Lorenz suggested a nonlinear chaotic model 
for an unspecified observable with next and second 
nearest neighbor couplings on grid points along
a latitude circle \cite{Lorenz:1996}.
Due to its scalability the model is 
a versatile tool in statistical mechanics
\cite{Abramov:2007,Hallerberg:2010,
Lucarini_JSP:2012,Lucarini_Sarno:2011}
and meteorology 
\cite{Ambadan:2009, Khare:2011, Messner:2011}.
The nonlinear terms have a quadratic conservation law 
and satisfy Liouville's Theorem.
For strong forcing the model shows  
intermittency \cite{Lorenz:2006}.



The Lorenz-96 equations for the variable $X_i$
are a surrogate for nonlinear advection  
in a periodic domain
\begin{equation} \label{dXdtLor96}
   \frac{d}{dt} X_i 
   = X_{i-1} [X_{i+1} - X_{i-2}]  
    - \gamma X_i + F_i
\end{equation}
$\gamma$ characterizes linear friction
($\gamma=1$ in \cite{Lorenz:1996}) and 
$F$ is a forcing.


In this Letter a continuous long wave approximation 
of the Lorenz-96 model is derived.
A surprising finding is that the 
nonlinear terms in the Taylor expansion 
are associated with generic dynamic operators. 
Furthermore, the dynamics in a truncated version
reveals avalanches, breather-like excitations
and flow reversals, which mimic various physical
processes in complex systems in a simplistic way.


Lorenz \cite{Lorenz:2005}
has analysed the linear stability of 
the mean $m$ of $X_i$ 
in (\ref{dXdtLor96}) and found that 
long waves with wave numbers $k<2 \pi/3$
are unstable for a positive mean $m$.


For $\gamma=F=0$ the equations (\ref{dXdtLor96}) 
are conservative with the
conservation law,
$H_X= 1/2 \sum_i X_i^2$,
denoted as energy in the following.
The dynamics in the state space of the $X_i$ is non-divergent
thus satisfying Liouville's Theorem,
$
 \sum_i \partial \dot{X}_i/\partial X_i =0
$.


The dynamics of an observable function $Q(X)$
is given by
   \begin{equation} \label{QQH}
   Q_t = \{Q,H_X\} 
   \end{equation}
with an anti-symmetric bracket 
\begin{equation} \label{ABJij}
   \{A,B\} 
   = \partial_i A J_{ij} 
   \partial_j B
   = -\{B,A\} 
\end{equation}    
and the antisymmetric matrix
\begin{equation} \label{JXij}
   J_{ij} = X_{i-1} \delta_{j,i+1} - X_{j-1} \delta_{i,j+1}
\end{equation}   
Energy $H_X$ is conserved due to the anti-symmetry of 
the bracket.

The conservative terms of the Lorenz-96 equations
(\ref{dXdtLor96}) are 
obtained for $Q=X_i$.
%
%
The equations are 
non-Hamiltonian \cite{Sergi:2001}
since the Jacobi identity 
\begin{equation} \label{JacobiX}
   \sum_\ell J_{i \ell} \frac{\partial J_{jk} }{\partial X_\ell} 
 + \sum_\ell J_{j \ell} \frac{\partial J_{ki} }{\partial X_\ell} 
 + \sum_\ell J_{k \ell} \frac{\partial J_{ij} }{\partial X_\ell} 
  = 0
\end{equation}
is not satisfied.


A continuous approximation is derived for 
a smooth dependency of $X_i$ on the spatial coordinate 
$x=ih$ in the limit $h \rightarrow 0$. 
%
%
The variable $X_i$ is replaced by a continuous function
$u(x,t)$ which is interpreted as velocity in the following.
We use the infinitesimal shift operators
\begin{equation}
L_{\pm} = \sum_{k=0}^{\infty} \frac{(\pm h \partial_{x})^{k}}{k!}
\end{equation}
to write the bracket (\ref{ABJij}) as 
\begin{equation}
	\lbrace A,B \rbrace 
	= \int \frac{\delta A}{\delta u} \mathcal{J}_{\infty} \frac{\delta B}{\delta u}
\end{equation}
%
with
\begin{equation}
	\mathcal{J}_{\infty} = (L_{-}u) \circ L_{+} - L_{-} \circ (L_{-}u)
\end{equation}
where $(L_{-}u)$ is a multiplication operator. 
The bracket is anti-symmetric since the adjoint is $L^{*}_{+} = L_{-}$.

By taking $n$-th order truncations of the operators $L_{\pm}$, 
we can find a hierarchy of truncated anti-symmetric operators 
\begin{eqnarray}
	\mathcal{J}_{nm} =  (L_{-,n} u) \circ L_{+,m} - L_{-,m} \circ (L_{-,n} u) 
\end{eqnarray}
where
\begin{equation}
	L_{\pm,n} = \sum_{k=0}^{n} \frac{(\pm h \partial_{x})^{k}}{k!}
\end{equation}
The total energy for the velocity $u(x,t)$ is
   \begin{equation} \label{HUcons}
   \mathcal{H} =    \frac{1}{2} \int u^2 dx
   \end{equation}
To each of these truncated operators corresponds a continuous 
Lorenz-96 model
\begin{equation} \label{utHnm}
	u_{t} = \lbrace u, \mathcal{H} \rbrace_{nm}
\end{equation}
where the indices indicate the operator $\mathcal{J}_{nm}$
(as in (\ref{dXdtLor96}) periodic boundary conditions are assumed).

The expansion of the nonlinear terms in (\ref{dXdtLor96})
up to order $O(h^2)$
yields for the rescaled coordinate 
$x^\prime = -x/3$ (the prime is dropped below)
  \begin{equation} \label{dudt2}   
     u_t = 
           -u u_x
     -
     \frac{1}{3} 
              \left(u_x^2+ \frac{1}{2} u u_{xx} \right)
     + f
    \end{equation}
%
%
%
with an advection and further nonlinear terms 
which are due to the noncentered definition
of the interaction in (\ref{dXdtLor96}).



%
The nonlinear terms 
are associated with 
antisymmetric evolution operators
    \begin{eqnarray} \label{ordersJ}
		O(h) &:&  
		\mathcal{J}_1 = -\frac{1}{3} (u \partial_x + \partial_x u)
		       \label{JOh1} \\
		O(h^2) &:&  
		\mathcal{J}_2 = -\frac{1}{6} (u_x \partial_x + \partial_x u_x)
		                          \label{JOh2} 
    \end{eqnarray}
Thus the evolution equation (\ref{dudt2})
can be written as
  \begin{equation} \label{utJHu}
   u_t = \mathcal{J} \frac{\delta \mathcal{H}}{\delta u},
   \qquad \mathcal{J}=\mathcal{J}_1+\mathcal{J}_2  
   \end{equation}
Note that the $O(h^3)$ expansion
in (\ref{utHnm}) is represented by a third operator 
$\mathcal{J}_3 = -(1/18) (u_{xx} \partial_x + \partial_x u_{xx})$;
here we restrict to the $O(h^2)$
expansion (\ref{dudt2}).


The evolution equation (\ref{dudt2})
has a conservation law 
   \begin{equation} \label{ConsLaw}
   \partial_t \left(\frac{1}{2}u^2 \right)
   = \partial_x \phi, 
   \end{equation}
   \begin{equation} \label{current}
   \phi = 
   -\frac{1}{3} u^3 
   - \frac{1}{6}  u^2 u_x
   \end{equation}
with the conserved current $\phi$
which leads to the conservation of total energy
(\ref{HUcons}).
Further conservation laws could not be found.
In particular momentum given as the mean flow
  \begin{equation} \label{umean}
   U = \langle u\rangle  
   = \int u dx
   \end{equation}
is not constant. 

In the following we consider a constant and positive forcing $f$
(note that the system is not dissipative).
In the presence of perturbations $v$ to the mean flow, $u=U+v$,
the mean flow energy $\bar{H}=U^2/2$ 
changes  according to
   \begin{equation} \label{dHqdt}
   \frac{\partial }{\partial t} \bar{H}
   = -\frac{U}{6} \langle v_x^2 \rangle + Uf
   \end{equation}
The perturbation energy 
\begin{equation} \label{TKE}
    E' = \frac{1}{2} \langle v^2 \rangle  
   \end{equation}
grows for positive $U$
   \begin{equation} \label{dTKEdt}
   \frac{\partial }{\partial t} E'
   = \frac{U}{6} \langle v_x^2 \rangle 
   \end{equation}
Thus, mean flows with $U>0$ ($U<0$) are unstable (stable)
as in the discrete system (\ref{dXdtLor96}) 
analysed in \cite{Lorenz:2005}.


The equations (\ref{dHqdt}, \ref{dTKEdt})
represent a coupling between perturbations and
the mean flow.
A forcing drives the mean flow   
towards positive values 
which allow the growth of perturbations.
When the perturbations are sufficiently intense
they reduce the flow to negative values
causing a decay of their intensities.




The nonlinear energy cycle represented by
the exchange between zonal flow and wave energy  
in (\ref{dHqdt}) and (\ref{dTKEdt})
is analysed in a spectral model
for the unstable long waves by Fourier expansion
in a periodic domain [$-\pi, \pi$]
  \begin{equation} \label{uexpans}
   u = \sum_{n=0}^N a_n \sin(nx) + b_n \cos (nx) 
   \end{equation}
Here we restrict to the low order system $N=2$.
    \begin{eqnarray} \label{danbndt}
		\dot{b}_0 &=& 
		           -\frac{1}{12}
		           \left(
		                 a_1^2
		               + b_1^2
		           \right)
		           -\frac{1}{3}
		             \left( a_2^2
		                   +b_2^2
		            \right)
		           +f
						                    \label{db0dt} \\
		\dot{a}_1 &=&
		            b_0 b_1
		            + \frac{1}{6}b_0 a_1 
		            + \frac{1}{2}
		            \left( a_1 a_2 + b_1 b_2 
		            \right)
		            \nonumber
		            \\
		            &+&
		            \frac{1}{4}
		            \left(  a_1 b_2 -b_1 a_2 
		            \right)
						                          \label{da1dt} \\
		\dot{b}_1 &=& 
		           -b_0 a_1 
		           +\frac{1}{6} b_0 b_1 
		           -\frac{1}{4}
		           \left(
		           a_1 a_2  +  b_1 b_2
		           \right)
		           \nonumber
		           \\
		           &+&\frac{1}{2}
		           \left(
                    a_1 b_2 
		              - b_1 a_2 
		           \right)
		 									         \label{db1dt} \\
		\dot{a}_2 &=&
		              2b_0 b_2
		              + \frac{2}{3} b_0 a_2
		              +\frac{1}{2}
		              \left(
		                  b_1^2 
		                 +a_1 b_1
		                 -a_1^2 
		              \right)
		                                              \label{da2dt} \\
		\dot{b}_2 &=& 
		             -2 b_0 a_2 
		             -a_1 b_1
		             +\frac{1}{4} 
		             \left( b_1^2 - a_1^2 
		             \right)
		             +\frac{2}{3} b_0 b_2 
		                                                \label{db2dt}
\end{eqnarray}

\begin{figure}[h]
\includegraphics[scale=0.8,angle=-90]{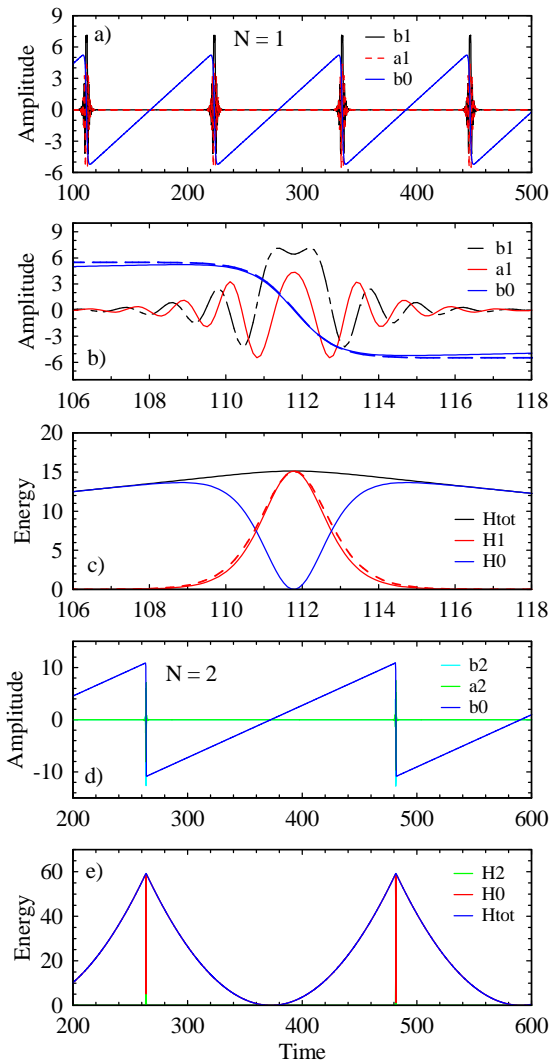}  
\caption{\label{fig1} 
Weak forcing $f=0.1$: 
(a) Amplitudes $b_0, a_1, b_1$ for $N=1$,
intervals $\approx 110$,
(b) amplitudes  
during a 
flow reversal with Eq. (\ref{b0N1})
for $b_0$ (dashed),  
(c) energies $H_0,H_1$ and Eq. (\ref{H1N1})
for $H_1$ (dashed),
(d) amplitudes $b_0, a_2, b_2$ for $N=2$,
($a_1, b_1$ vanish),
and
(e) energies ($H_1$ vanishes).
}
\end{figure}

The mean flow is $U= b_0$
which is subject to a constant forcing $f$ in 
the numerical experiments  (\ref{db0dt}).
The truncated system conserves energy 
  \begin{equation} \label{Htot012}
   H_{tot} =  H_0+H_1+H_2
   \end{equation}
   \begin{equation} \label{H0H1H2}
   H_0 = \frac{1}{2} b_0^2, 
   \quad 
   H_1 = \frac{1}{4} \left( a_1^2 + b_1^2 \right),
   \quad
   H_2 = \frac{1}{4} \left( a_2^2 + b_2^2 \right)
   \end{equation}
The Liouville Theorem is not satisfied 
  \begin{equation} \label{QTH}
   \sum_{n=0}^2 \left(
   \frac{\partial \dot{a}_n}{\partial a_n} 
   + \frac{\partial \dot{b}_n}{\partial b_n}
   \right)
   = \frac{5}{3}b_0
   \end{equation}
The expansion and contraction
of the state space volume is controlled
by the sign of the mean flow.



Numerical experiments reveal a flow reversal mechanism
and vanishing long term means
of mean flow and wave number amplitudes,
hence the Liouville Theorem (\ref{QTH}) 
is satisfied in the mean.  


(i) Weak forcing with $f=0.1$ in the $N=1$ truncation   
reveals periodic flow reversals
(Fig. \ref{fig1}).
The system starts with 
randomly chosen amplitudes.
A mean flow increases gradually to positive 
values where it becomes unstable due to the excited 
waves, denoted as breathers in the following. 
These breathers drive a rapid flow reversal 
towards a negative flow which initiates their collapse. 
The process is energy conserving 
on short time scales.
The total energy increases (decreases) 
when the mean flow is positive (negative). 

For $N=1$ with the amplitudes $b_0, a_1, b_1$ 
the energy cycle 
is for $f=0$ (compare (\ref{dHqdt}, \ref{dTKEdt})) 
   \begin{equation} \label{H0H1}
      \partial_t H_0 = -\frac{1}{3} b_0 H_1, \qquad
      \partial_t H_1 = \frac{1}{3} b_0 H_1
   \end{equation}
which is controlled by the mean flow.  
The solution for the mean flow is for $b_0(0)=0$
  \begin{equation} \label{b0N1}
     b_0 
     = -6a \tanh(at)     
   \end{equation}
and the perturbation energy is 
  \begin{equation} \label{H1N1}
     H_1 
     =\frac{18a^2}{\cosh^2(at)}
     \end{equation}
where $a$ is related to the total energy $H=18a^2$.
$H_1$ attains its maximum during flow reversals when $U=0$.
These approximations are compared to the forced simulation
in Fig. \ref{fig1}b,c centered at a single flow reversal.

In the presence of forcing $f$ and for a
small wave energy $H_1$ the mean flow  $b_0$ grows 
linearly in time, $b_0(t) \approx ft$, 
up to a value $b_{0,max}$.
This defines an interarrival time scale
of flow reversals, $\tau=2 b_{0,max}/f$.
In this range the wave energy evolves rapidly according to
$H_1(t) \sim \exp(ft^2/6)$. 


\begin{figure}[h]
\includegraphics[scale=0.8,angle=-90]{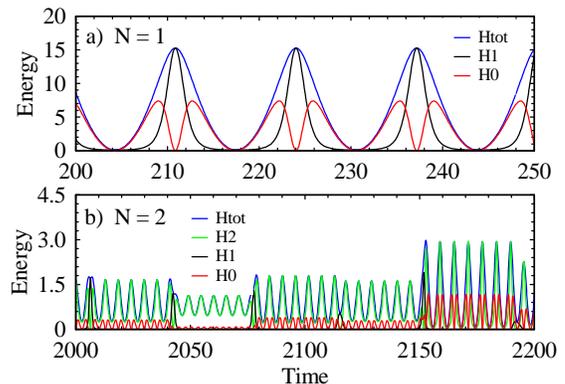}  
\caption{\label{fig2} 
Intermediate forcing $f=1$: Energy distributions for 
(a) $N=1$, intervals $\approx 13$,
(b) $N=2$.
}
\end{figure}

The described flow reversal mechanism is retained for 
viscous dissipation represented by a linear damping 
of the wave amplitudes $a_1$ and $b_1$.

For the $N=2$ truncation with all modes
$b_0, a_1, b_1, a_2$ and $b_2$ 
flow reversals occur on a time scale
roughly twice as for $N=1$ (Fig. \ref{fig1}d).
Due to the weak forcing the energy cascades 
to mode 2 with negligible amplitudes 
$a_1, b_1$ and energy $H_1$ (Fig. \ref{fig1}d, e).
Neglecting the modes 1, the energy cycle for 
interactions among $b_0, a_2$ and $b_2$ is
   \begin{equation} \label{H0H2}
      \partial_t H_0 = -\frac{4}{3} b_0 H_2, \qquad
      \partial_t H_2 = \frac{4}{3} b_0 H_2
   \end{equation}
This corresponds to a rescaling of the $H_0 - H_1$
cycle (\ref{H0H1}) by $\tilde{t}=2t$ for time 
and $\tilde{b}_0=2b_0$ etc. for the amplitudes,
hence the energies quadruple.

\begin{figure}[h]
\includegraphics[scale=0.8,angle=-90]{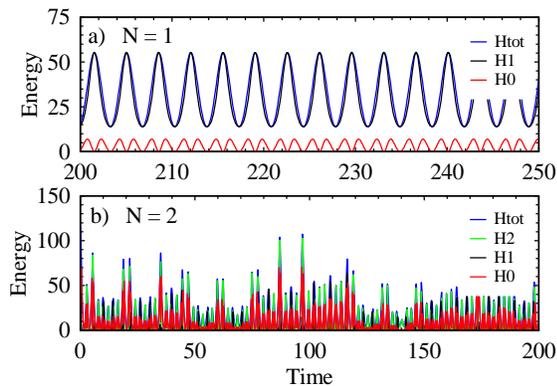}  
\caption{\label{fig3} 
Strong forcing $f=10$: energy distributions for 
(a) $N=1$, intervals $\approx 3$,
(b) $N=2$.
}
\end{figure}


(ii) For intermediate forcing with $f=1$ 
the time scale between flow reversals decreases 
by an order of magnitude in the $N=1$ truncation 
(see Fig. \ref{fig2}a).
Thus the intervals $\tau$ approach the duration of 
individual breathers.
For the complete set of modes in $N=2$ 
(Fig. \ref{fig2}b) 
the system is weakly nonlinear 
with a mixing of frequencies, 
$\omega/2, \omega, 3\omega/2$ and $2 \omega$,
where $\omega=2 \pi /\tau$ is defined by the interarrival times of
the flow reversals \cite{Lucarini:2009}.
The lowest frequency determines the amplitude modulation.


(iii) For strong forcing, $f=10$, the 
flow reversals in the $N=1$ truncation
are regular (Fig. \ref{fig3}a) with
intervals decreased by an order of magnitude 
relative to $f=1$.
The dominant part of energy is accumulated in waves.
In the $N=2$ truncation the dynamics becomes 
intermittent as in the regime behaviour
detected by Lorenz \cite{Lorenz:2006} 
in the discrete equations
(\ref{dXdtLor96}).
The events loose their identities and
the systen becomes strongly nonlinear.



In Summary, a continuous dynamical equation derived 
from the Lorenz-96 model
is able to mimic 
several types of complex processes observed 
in geophysics, geophysical fluid dynamics,
and solid state physics:

(i) Avalanche processes excited by continuous
driving as in the sand pile model of Bak et al. 
\cite{Bak_SOC:1987};
see also the recent observation
of quasi-periodic events in crystal plasticity
subject to external stress \cite{Papanikolaou:2012}.
A common characteristic property is the weakness of 
the external forcing 
which is necessary to cause avalanches.
In the present model the flow is driven by a constant
forcig towards a state where
mean flow and wave energy interact.
The intervals between the flow reversals  are
approximately proportional to the inverse of
the forcing intensity, $\sim 1/f$.

(ii) The Quasi-Biennial Oscillation (QBO, 
\cite{Baldwin:2001}),
a flow reversal in the tropical stratosphere driven
by two different types of upward propagating gravity waves.
A common aspect is that the driving 
of the mean flow by waves occurs only for
a particular sign of the mean flow.
Although the QBO is considered to be explained 
dynamically the simulation in present-day weather 
and climate models necessitates
careful sub-scale parameterizations
or high resolution models \cite{Kawatani:2010}.
The present model is clearly an oversimplification
but can be considered as a toy model for this phenomenon.

(iii) Rogue waves (also termed freak or monster waves) 
at the ocean surface are simulated mainly 
by the nonlinear Schroedinger equation 
(e.g. \cite{Onorato:2006,Calini:2012});  
a Lagrangian analysis has been published 
recently \cite{Abrashkin:2013}.
The breather solutions found in the present model
show characteristics like the rapid evolution
and the high intensity in an almost quiescent medium.

Due to the flow reversals the total energy 
of the non-dissipative system remains finite 
for a constant forcing.   
The long term mean of the mean flow vanishes 
and the Liouville Theorem (\ref{QTH}) 
is satisfied in the mean.  
The flow reversals are insensitive to viscous dissipation.





\begin{acknowledgments}
JW and VL acknowledge support from the 
European Research Council under the European Community’s
Seventh Framework Programme (FP7/2007-2013)/ERC Grant
agreement No. 257106. 
We like to thank the cluster of excellence clisap at 
the University of Hamburg.
\end{acknowledgments}

\bibliography{Lorenz}

\end{document}